\documentclass[aps,prb,twocolumn,showpacs,amsmath]{revtex4}
\usepackage{graphicx}
\usepackage{bm}

\begin{document}

\title{Field and current distributions and ac losses in superconducting strips}

\author{John R.\ Clem}
\affiliation{%
   Ames Laboratory and Department of Physics and Astronomy, \\
   Iowa State University, Ames, Iowa, 50011--3160}

\date{\today}

\begin{abstract} 
In this paper I discuss analytic and numerical calculations of the magnetic-field and sheet-current distributions in superconducting strips of width $2a$ and arbitrary thickness $2b$ at the center when the cross section is an ellipse, a rectangle, and a shape intermediate between these limits.   
Using critical-state theory, I use several methods to determine the functional dependence of the ac transport-current losses upon $F = I/I_c$, where $I$ is the peak alternating current and $I_c$ is the critical current, and I discuss how this dependence can be affected by the cross-sectional shape, aspect ratio, and a flux-density-dependent critical current density $J_c(B)$.
\end{abstract}

%PACS 2006
%74.25.Ha 	Magnetic properties
%74.25.Nf 	Response to electromagnetic fields (nuclear magnetic resonance,
%surface impedance, etc.)
%74.25.Op 	Mixed states, critical fields, and surface sheaths
%74.25.Sv 	Critical currents
%74.78.Bz 	High-Tc films
\maketitle

\section{\label{intro}Introduction} %***** 

In determining the usefulness of a type-II superconductor in applications, the ac losses are a very important factor.  Many composite conductors currently under development for large-scale applications of superconductivity are in the form of tapes, consisting of superconducting strips embedded in a nonsuperconducting metallic matrix.  In this paper I focus on the self-field hysteretic ac losses of such superconducting strips carrying an ac current, neglecting the possibility of losses in the surrounding matrix.  

In a classic paper, Norris\cite{Norris70} investigated the hysteretic ac losses in type-II superconductors with a variety of cross sections, and he derived results for the ac transport-current losses expressed in powers of $F = I/I_c$, where $I$ is the peak alternating current and $I_c$ is the critical current.  For small values of $F$, Norris found that the losses were proportional to $F^3$ for wires of elliptical or circular cross section but were proportional to $F^4$ for thin strips of rectangular cross section.  The large difference in the power-law behavior seems puzzling in view of the fact that films of elliptical cross section look very similar to films of rectangular cross section when both films are thin. 

To analyze the reasons for this difference in power-law behavior, there are at least three theoretical questions that need to be addressed.  First, how thin must a rectangular strip of width $2a$ and thickness $2b$ be in order for the ac losses to be well described by the thin-film limit studied by Norris? Second, assuming that a film of width $2a$ and thickness $2b$ in the middle is thin enough to be described by the thin-film limit, how do the ac losses depend upon the cross-sectional shape if the cross section is neither a rectangle nor an ellipse but something in between, as shown in Fig. 1?
Third, since the critical current density $J_c$ depends in general upon the local magnetic flux density $B$, does the $B$ dependence of $J_c$ have a significant effect?

\begin{figure}%***** Fig.1 ************************
\includegraphics[width=8cm]{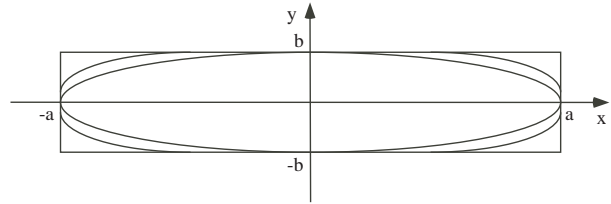}
\caption{Cross sections considered in this paper, all representing superconducting strips of width $2a$ and thickness $d=2b$ at the center: an ellipse of semimajor axis $a$ and semiminor axis $b$, a rectangle of length $2a$ and height $2b$, and an intermediate shape, described in the text. }
\label{Fig01}
\end{figure}

Experimental questions regarding the power-law behavior also have been raised.
Some measurements\cite{Kerchner97, Miyagi01} of the ac transport-current ac losses in YBCO films and Bi-2223 tapes have been found to deviate from the  $F^4$ behavior expected for thin strips of rectangular cross section.  

In this paper I theoretically explore how the transport ac losses of a superconducting strip depend upon $F = I/I_c$, and I discuss how this functional dependence can be affected by the superconductor's cross-sectional shape, aspect ratio, and $J_c(B)$.  
I begin in Sec.\ \ref{fieldcurrent} by discussing ways to calculate the field and current distributions in the critical state.  In Sec.\ \ref{fieldthick} I discuss thick strips with cross sections described by a shape function $y_c(x)$, and I model the inner boundary of the flux-penetrated region by a similar function $y_I(x)$.  In Sec.\ \ref{fieldthin} I present results for the field and current distributions for superconducting strips of different cross-sectional shapes $y_c(x)$ in the thin-film limit.  In Sec.\ \ref{losses} I apply these results to calculate the hysteretic ac transport  losses.  After discussing some general methods in Sec.\ \ref{lossesmethods}, I discuss the losses in conductors with elliptical cross section in Sec.\ \ref{losseselliptical} and with rectangular cross section in Sec.\ \ref{lossesrectangular}.  For the thin-film limit I calculate the losses in conductors with elliptical, rectangular, and intermediate cross sections in Sec.\ \ref{lossesthin}, and I show how to account approximately for the $B$ dependence of $J_c(B)$ in Sec.\  \ref{Jc(B)}.  Finally, I briefly summarize my results in Sec.\ \ref{discussion}.

\section{Field and current distributions\label{fieldcurrent}}

\subsection{Thick strips\label{fieldthick}}

To simplify the calculations and obtain analytic results,  Norris used critical-state theory,\cite{Campbell72} assuming that  the relation between the magnetic induction $\bm B$ and the magnetic field $\bm H$ is ${\bm B} = \mu_0 {\bm H}$ and that the critical current density $J_c$ is independent of $B$.  I use the same assumptions here.   
Norris noted that the ac loss calculation is greatly simplified by first finding the shape of the cross-sectional area of the flux front at the peak current $I$, and he found that for superconductors of elliptical cross section the flux front is an ellipse with the same aspect ratio as that of the superconductor itself.  Although Norris did not report the shape of the flux front for very thin superconductors of rectangular cross section, the shape can be inferred from the current and field distributions.\cite{Brandt93,Zeldov94}
In this paper, I assume a function with two fitting parameters to approximate the flux-front shape.  Although this shape function does not yield exact results, the resulting magnetic field and current distributions are very nearly correct.  I then use these distributions to calculate the ac losses.

Sketched in Fig.\ 1 are the sample cross sections to be considered in this paper.  The boundary surface can be described by  $y = \pm y_c(x)$, where
\begin{equation}
y_c(x)=\frac{b \tan^{-1}\!\!\sqrt{\alpha^2(1-x^2/a^2)}}{\tan^{-1}\!\alpha}
\label{y(x)}
\end{equation}
or its inverse, $x = \pm x_c(y)$, where
\begin{equation}
x_c(y) = a\sqrt{1-\frac{1}{\alpha^2}\tan^2\Big(\frac{y\tan^{-1}\!\alpha}{b}\Big)}
\label{x(y)}
\end{equation}
and $\alpha$ is a dimensionless parameter that can range from 0 to $\infty$.  In the limit $\alpha \to 0$, these equations describe an ellipse of semiaxes $a$ and $b$; in the limit $\alpha \to \infty$, they describe a rectangle of dimensions $2a \times 2b$; and for an intermediate value of $\alpha$, they describe the intermediate shape shown in Fig.\ 1.

The radius of curvature of the curve $y_c(x)$ at $x = a$ or of $x_c(y)$ at $y = 0$ is 
\begin{equation}
R_c = \Big(\frac{\alpha}{\tan^{-1}\alpha}\Big)^2 \frac{b^2}{a}.
\label{Ra}
\end{equation}
For reference, note that $R_c = b$ when $\alpha = 0$ for $b/a= 1$,  $\alpha = 4.23$ for $b/a= 0.1$, $\alpha = 15.0$ for $b/a=0.01$, $\alpha = 49.0$ for $b/a= 0.001$, $\alpha = 156$ for $b/a= 0.0001$, and $\alpha = 496$ for $b/a= 0.00001$. 

The cross-sectional area $S_c$ of the sample is
\begin{equation}
S_c=\frac{2\pi a b (\sqrt{1+\alpha^2}-1)}{\alpha \tan^{-1}\!\alpha},
\label{Sc}
\end{equation}
such that $S_c \to \pi a b$ when $\alpha \to 0$; $S_c \to 4 ab$ when $\alpha \to \infty$; and $\pi a b < S_c < 4ab$ when $ 0 < \alpha < \infty.$

When a uniform current density $J_z = J_c$ is flowing throughout the entire cross section $S_c$, the vector potential is ${\bm A}_c(x,y) =A_{cz}(x,y) {\hat z}, $
\begin{equation}
A_{cz}(x,y)\!=\!-\frac{\mu_0 J_c}{4\pi}\!\!\!\int \!\!\!\!\int_{S_c}   \! \!du dv \log\Big[\frac{(x\!-\!u)^2\!+\!(y\!-\!v)^2}{u^2+v^2}\Big],
\label{Acz}
\end{equation}
where the integral over $u$ and $v$ extends over the area $S_c$ and the constant of integration has been chosen such that $A_{cz}(0,0)= 0$.  Expressions for $A_{cze}(x,y)$ for the elliptical cross section and $A_{czr}(x,y)$ for the rectangular cross section are given in Appendixes A and B, but $A_{czi}(x,y)$ for the intermediate case is most conveniently obtained by numerical integration using upper and lower limits obtained from Eq.\ (\ref{y(x)}) or (\ref{x(y)}). The magnetic induction is  ${\bm B}_c(x,y) = \mu_0 {\bm H}_c(x,y) = \nabla \times {\bm A}_{c}(x,y)$. 

On the other hand, when a current $I < I_c$ is applied in the $z$ direction to a sample originally in the virgin state containing no magnetic flux, magnetic flux pushes its way into the sample.  The leading edge of the flux front encircles an area $S_I$, here approximated as a roughly elliptical shape of width $2c$ and height $2y_0$, with a boundary in the first quadrant given by equations of the same form as Eqs.\ (\ref{y(x)}) and (\ref{x(y)}):
\begin{equation}
y_I(x)=\frac{y_0 \tan^{-1}\!\!\sqrt{\beta^2(1-x^2/c^2)}}{\tan^{-1}\!\beta}
\label{yI(x)}
\end{equation}
or its inverse,
\begin{equation}
x_I(y) =c\sqrt{1-\frac{1}{\beta^2}\tan^2\Big(\frac{y\tan^{-1}\!\beta}{y_0}\Big)},
\label{xI(y)}
\end{equation}
where $\beta$ is a dimensionless parameter that can range from 0 to $\infty$.  The area enclosed by the flux front is
\begin{equation}
S_I=\frac{2\pi c y_0 (\sqrt{1+\beta^2}-1)}{\beta \tan^{-1}\!\beta},
\label{SI}
\end{equation}

Consider, as an auxiliary function, the vector potential ${\bm A}_I(x,y) = A_{Iz}(x,y){\hat z}$ generated by a uniform current density $J_z = -J_c$  flowing only in the cross section $S_I$,
\begin{equation}
A_{Iz}(x,y)\!=\!\frac{\mu_0 J_c}{4\pi}\!\!\!\int \!\!\!\!\int_{S_I}   \! \!du dv \log\Big[\frac{(x\!-\!u)^2\!+\!(y\!-\!v)^2}{u^2+v^2}\Big],
\label{AIz}
\end{equation}
where the integral over $u$ and $v$ extends over the area $S_I$, and the constant of integration again has been chosen such that $A_{Iz}(0,0)= 0$.  
The corresponding magnetic induction is  ${\bm B}_I(x,y) = \mu_0 {\bm H}_I(x,y) = \nabla \times {\bm A}_{I}(x,y)$. 

When the current $I<I_c$ is applied in the $z$ direction, the current density is $J_z= J_c$ in the area $S_p = S_c-S_I$, the flux-penetrated portion of  $S_c$ outside the area $S_I$, and $J_z=0$ inside the area $S_I$.  The resulting vector potential ${\bm A}(x,y) =A_z(x,y) {\hat z}$ is 
\begin{eqnarray}
A_z(x,y)\!&=&\!A_{cz}(x,y)+A_{Iz}(x,y),\nonumber \\
\!&=&\!\frac{\mu_0 J_c}{4\pi}\!\!\!\int \!\!\!\!\int_{S_p}   \! \!du dv \log\Big[\frac{(x\!-\!u)^2\!+\!(y\!-\!v)^2}{u^2+v^2}\Big],
\label{Az}
\end{eqnarray}
subject to the condition that the shape of the area $S_I$ is such that $A_z=0$ inside the area $S_I$, and the corresponding magnetic flux density 
\begin{equation}
{\bm B}(x,y) = {\bm B}_c(x,y) +{\bm B}_I(x,y)
\label{B}
\end{equation}
is also zero there.  The reduced current $F=I/I_c$, where $I_c = J_c S_c$, obeys
\begin{equation}
F=S_p/S_c.
\label{FvsSp}
\end{equation}

Norris\cite{Norris70} showed that if the cross section of the superconductor is an ellipse of semiaxes $a$ and $b$, such that the bounding surface is described by Eqs.\ (\ref{y(x)}) and (\ref{x(y)}) with $\alpha \to 0$ and $S_c = \pi a b$, the flux front encloses the area $S_I = \pi c y_0$, an ellipse of semiaxes $c$ and $y_0$ described by Eqs.\ (\ref{yI(x)}) and (\ref{xI(y)}) with $\beta \to 0$, where $y_0/c = b/a$.  The reduced current $F=I/I_c$ obeys $F = 1-c^2/a^2$.

Norris\cite{Norris70} also considered a flat superconducting strip of width $2a$ and thickness $2b \ll 2a$, and obtained the current density averaged over the thickness, which can be written as
\begin{eqnarray}
\bar J_z(x) &=& \frac{2 J_c}{\pi} \tan^{-1}\sqrt{\frac{a^2-c^2}{c^2-x^2}},\; |x|<c,
\\
&=& J_c,\; c \le |x| \le a,
\label{NorrisJz}
\end{eqnarray}
where  $F=I/I_c = \sqrt{1-c^2/a^2}$ is the reduced current.
Since $\bar J_z(x) = J_c[1-y_I(x)/b]$, the area $S_I$ has only a roughly elliptical shape of width $2c$ with upper and lower boundaries at $y = \pm y_I(x)$ as given in Eq.\ (\ref{yI(x)}), where $\beta^2=c^2/(a^2-c^2)=(1-F^2)/F^2$ and $y_0 = (2b/\pi)\tan^{-1}\!\beta.$  When $F \to 0$, $\beta \to \infty$, and $S_I$ becomes a rectangle with width $2a$ and height $2b$, filling the entire cross section.  When $F \to 1$, $\beta \to 0$, and $S_I$ becomes a small ellipse with semimajor axis $c \approx a\sqrt{1-F^2}$ and semiminor axis $y_0 \approx (2b/\pi)\sqrt{1-F^2}$, so that $y_0/c = (\pi/2)b/a$.

To obtain approximate results for the vector potential $A_z(x,y)\hat z$  and the corresponding magnetic flux density ${\bm B}(x,y)$ for sample cross sections that are intermediate between an ellipse and a very thin flat strip, we can use the following procedure.  When the current $I < I_c$, we assume that the vector potential is given by Eq.\ (\ref{Az}), where the auxiliary vector potential 
$A_{Iz}(x,y)$ depends upon the shape $S_I$, which in turn is characterized by three fitting parameters, $c$, $y_0$, and $\beta$.  For a given value of $F = I/I_c$, we can determine these parameters from three equations, Eq.\ (\ref{FvsSp}), $B_y(c,0)=0$, and $B_x(0,y_0)=0$.  

For a rectangular cross section we can use Eq.\ (\ref{B}), evaluate ${\bm B}_c(x,y)$ analytically using Appendix B, and calculate ${\bm B}_I(x,y)$ numerically using Appendix C. Examples of the results of this procedure are shown in Fig.\ 2, which exhibits plots of $B_y(x,0)$ and $B_x(0,y)$ vs $x$ for $b/a$ = 1/2 and a series of values of $c/a$. When $b/a$ is not very small (as in the case shown), $B_y$ and $B_x$ vary nearly linearly with distance near the sample surface.  Table \ref{1}  exhibits the corresponding values of $y_0/a$, $\beta$, $S_I/S_c$, and $F = I/I_c$.  The solid curves in Fig.\ \ref{Fig03} show plots of $\Delta_x = (a-c)/a$, calculated as above, vs $F = I/I_c$ for a variety of values of $b/a$. For  $F \ll 1$ and modest values of $b/a$, $\Delta_x \propto F;$ note that $\Delta_x \propto F^2$ only for very small values of $b/a$. 

\begin{figure}%***** Fig.2 ************************
\includegraphics[width=8cm]{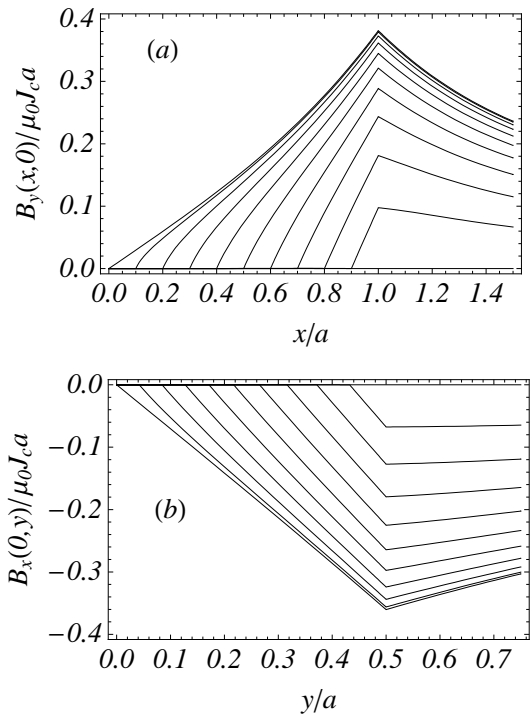}
\caption{Plots of (a) $B_y(x,0)/\mu_0J_ca$ and (b) $B_x(0,y)/\mu_0J_ca$ vs $x/a$ for $c/a$ = 0, 0.1, 0.2, 0.3, 0.4, 0.5, 0.6, 0.7, 0.8, 0.9, and 1.0 in a strip of rectangular cross section of dimensions $2a \times 2b$, where $b = a/2$. The corresponding values of $y_0$, $\beta$, $S_I/S_c$, and $F = I/ I_c$ are given in Table I.}
\label{Fig02}
\end{figure}

\begin{table}
\centering
\caption{\label{1} Fitting parameters for $S_I$ ($c$, $y_0$, and $\beta$), $S_I/S_c$, and $F = I/I_c$ for the plots shown in Fig.\ 2 obtained from the requirements that $B_x(0,y_0)=0$ and $B_y(c,0)=0$.  The strip has a rectangular cross section of dimensions  $2a \times 2b$, where $b = a/2$, such that $S_c = 2a^2.$}
\begin{ruledtabular}
\begin{tabular}{ccccc}
{$c/a$} & {$y_0/a$} & {$\beta$} & $S_I/S_c$ & {$F$} \\ 
\hline
0.0 & 0.000 & 0.000 & 0.000 & 1.000 \\
0.1 & 0.042 & 0.141 & 0.007 & 0.993\\
0.2 & 0.084 & 0.290 & 0.027 & 0.973 \\
0.3 & 0.127 & 0.453 & 0.061 & 0.939 \\
0.4 & 0.171 & 0.642 & 0.111 & 0.889 \\
0.5 & 0.217 & 0.878 & 0.178 & 0.822\\
0.6 & 0.265 & 1.197 & 0.267 & 0.733 \\
0.7 & 0.316 & 1.685 & 0.382 & 0.618 \\
0.8 & 0.371 & 2.596 & 0.533 & 0.467 \\
0.9 & 0.432 & 5.184 & 0.731 & 0.269 \\
1.0 & 0.500 & $\infty$ & 1.000 & 0.000
\end{tabular}
\end{ruledtabular}
\end{table}

\begin{figure}%***** Fig.3 ************************
\includegraphics[width=8cm]{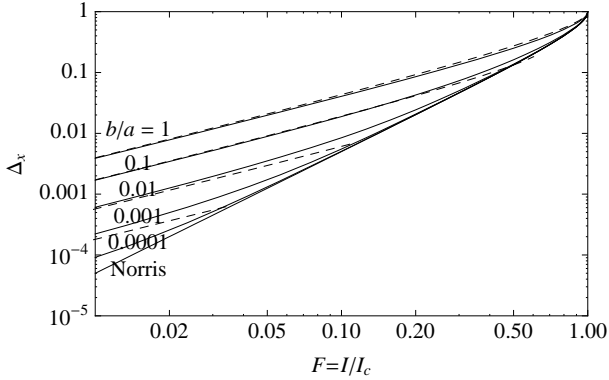}
\caption{The upper five solid curves show values of $\Delta_x = (a-c)/a$ vs $F = I/I_c$ for a strip of rectangular cross section ($2a \times 2b$), numerically calculated as described in the text for $b/a$ = 1, 0.1, 0.01, 0.001, and 0.0001. The corresponding dashed curves show results using the conformal-mapping method described in Sec.\ \ref{lossesrectangularconformal}. The lowest solid curve shows  $\Delta_x = 1-\sqrt{1-F^2}$, Norris's result for a very thin strip of rectangular cross section.\cite{Norris70} }
\label{Fig03}
\end{figure}

\begin{figure}%***** Fig.4 ************************
\includegraphics[width=8cm]{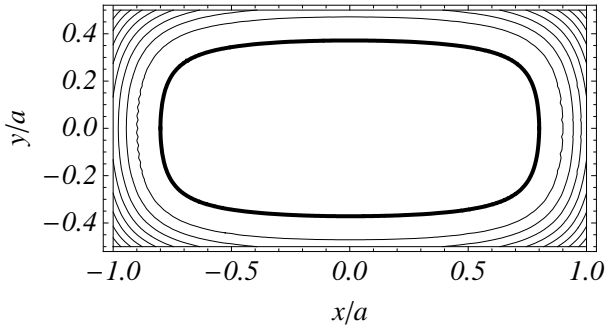}
\caption{Contour plot of $A_z(x,y)$ vs $x/a$ and $y/a$ for $c/a$ =  0.8, $y_0/a$ = 0.372, $\beta =$ 2.629, $S_I/S_c = 0.534$, and $F = I/I_c$ = 0.466 in a strip of rectangular cross section of dimensions $2a \times 2b$, where $b = a/2$.  The flux front, described by $y=\pm y_I(x)$ [Eq.\ (\ref{yI(x)})] and shown as the bold curve, surrounds the area $S_I$. The contours correspond to magnetic field lines, which circulate around $S_I$ in the counterclockwise direction. }
\label{Fig04}
\end{figure}

Figure \ref{Fig04} shows a contour plot of $A_z(x,y)$ vs $x$ and $y$,  calculated as described above, for a strip of rectangular cross section.  The contours correspond to magnetic field lines, which in principle do not penetrate into the area $S_I$ bounded by the bold curve.  However, the assumed shape of the area $S_I$ as approximated by $y = \pm y_I(x)$ [Eq.\ (\ref{yI(x)})] or $x = \pm x_I(y)$ [Eq.\ (\ref{xI(y)})] does not give exact solutions for either the vector potential $A_z(x,y)$ or the magnetic induction ${\bm B}(x,y) = \mu_0 {\bm H}(x,y)$.  Nevertheless, for the cases shown in Figs.\ \ref{Fig02} and \ref{Fig04}, the calculated values of  $B(x,y) = \sqrt{B_y^2(x,y) + B_x^2(x,y)}$ and $A_z(x,y)$ inside the area $S_I$, though not precisely equal to zero, are about three or four orders of magnitude smaller than  their values on the perimeter of the strip.  These results indicate that the values of  $A_z(x,y)$ or the magnetic induction ${\bm B}(x,y) = \mu_0 {\bm H}(x,y)$ calculated as above are reasonable approximations to the exact solutions, thereby permitting relatively simple calculations showing  how the hysteretic ac losses depend upon the cross-sectional shape.

\subsection{Thin-film limit\label{fieldthin}}

The magnetic fields generated by long thin strips can be calculated as in Refs.\ \onlinecite{Norris70}, 
\onlinecite{Brandt93}, and \onlinecite{Zeldov94}, using the method of complex fields.  Instead of dealing with the vector field ${\bm H}(x,y) = {\bm B}(x,y)/\mu_0 =\hat x H_x(x,y) + \hat y H_y(x,y)$, one works with the complex field ${\cal H}(\zeta) = H_y +i H_x$, which is an analytic function of $\zeta = x+iy$ outside the strip.  Since analytic functions obey the Cauchy relations, the conditions that $\nabla \cdot \bm H = 0$ and $\nabla \times \bm H = 0$ are automatically satisfied.  

In the limit as $b/a \to 0$, the complex magnetic field outside a thin strip whose boundary surface is described by Eq.\ (\ref{y(x)}) and which carries a total current $I$ with a current density $J_z = J_c$  at the edges ($c \le |x| < a$) and  an average current density $\bar J_z(x) < J_c$ in the middle  ($|x| < c$)  is
\begin{eqnarray}
{\cal H}(\zeta) &=&\frac{J_c b}{\tan^{-1}\alpha}\Big[\tanh^{-1}\Big(\frac{\alpha \sqrt{\tilde \zeta^2-\tilde c^2}}{\sqrt{1+\alpha ^2(1-\tilde c^2)}}\Big)
\nonumber \\
&&-\tanh^{-1}\Big(\alpha \sqrt{\tilde \zeta^2 -1}\Big) \Big],
\label{complexH}
\end{eqnarray}
where $\tilde \zeta = \zeta/a$ and $\tilde c = c/a$.  The condition relating $I$ and $c$ is
\begin{equation}
F = \frac{I}{I_c}=\frac{\sqrt{1+\alpha ^2(1-\tilde c^2)}-1}{\sqrt{1+\alpha ^2} -1},
\label{Fvsc}
\end{equation}
where $I_c=J_c S_c$ [see Eq.\ (\ref{Sc})].  Here the notation $\sqrt{\tilde \zeta^2-\tilde c^2}$ is an abbreviation for $(\tilde \zeta-\tilde c)^{1/2}(\tilde \zeta+\tilde c)^{1/2}$.

Equations (\ref{complexH}) and (\ref{Fvsc}) reduce to
\begin{eqnarray}
{\cal H}(\zeta) &=&J_c b \Big(\sqrt{\tilde \zeta^2-\tilde c^2}-\sqrt{\tilde \zeta^2 -1}\Big),\\
F&=&1-\tilde c^2,
\label{F0}
\end{eqnarray}
for an elliptical cross section ($\alpha = 0$) and to
\begin{eqnarray}
{\cal H}(\zeta) &=&\frac{2 J_c b}{\pi}\tanh^{-1}\sqrt{\frac{1-\tilde c^2}{\tilde \zeta^2-\tilde c^2}},\\
F&=&\sqrt{1-\tilde c^2},
\label{Finfty}
\end{eqnarray}
for a rectangular cross section ($\alpha = \infty$).

\begin{figure}%***** Fig.5 ************************
\includegraphics[width=8cm]{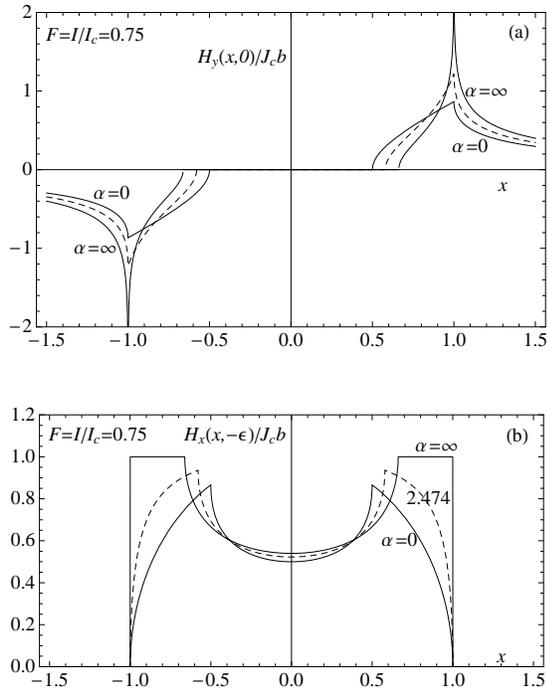}
\caption{Plots of $H_y(x,0)$ and $H_x(x,-\epsilon)$ (on the bottom of the strip) calculated from Eqs.\ (\ref{complexH})-(\ref{Finfty}) for $a = 1$, $\alpha$ = 0, 2.474 (dashed), and $\infty$, and $F=I/I_c = 0.75$, for which $c/a$ = 0.500, 0.579, and 0.661.}
\label{Fig05}
\end{figure} 

Shown in Fig.\ \ref{Fig05} are plots of  $H_y(x,0)$ and $H_x(x,-\epsilon)$  obtained from Eqs.\ (\ref{complexH})-(\ref{Finfty}) for $F = I/I_c = 0.75$ and $\alpha$ = 0 (solid curves), 2.474 (dashed), and $\infty$ (solid).  The cross-sectional area for $\alpha$ = 2.474 is 3.57$ab$, the average of the areas of an ellipse ($\pi ab$) and a rectangle ($4ab$).  In the regions $c \le |x| < a$, $J_z = J_c$ and $H_x(x,-\epsilon) = J_c y_c(x)$ [see Eq.\ (\ref{y(x)})].

\section{ac losses\label{losses}}

\subsection{Methods\label{lossesmethods}}

We are now in a position to analyze some general features of the hysteretic transport ac losses in isolated  superconducting strips. Let us consider ac currents of amplitude $I$ less than $I_c$ at frequencies $f = 1/T$ that are sufficiently low that eddy-current losses are negligible and the losses can be calculated using a quasistatic approach.\cite{Norris70}  The solutions for ${\bm B}(x,y)=\mu_0{\bm H}(x,y)$ derived in Sec.\ II 
can be used to calculate $Q'$, the energy dissipated per cycle per unit length.  Consider time $t=0$, when the current has its maximum value  $I$ in the $z$ direction, the magnetic-field distribution is given by ${\bm H}(x,y) = \hat x H_x(x,y) + \hat y H_y(x,y)$, the magnetic induction is ${\bm B}(x,y) = \mu_0 {\bm H}(x,y)$, and the vector potential is ${\bm A}(x,y)=A_z(x,y) {\hat z}$.  Half a cycle earlier, at time $t = -T/2$, when the current was in the opposite direction, ${\bm B}(x,y,-T/2) = -\mu_0 {\bm H}(x,y)$. The loss per cycle per unit length $Q'$ is twice the loss in the half cycle $-T/2 \le t \le 0.$ Thus 
\begin{equation}
Q' = 2\int_{-T/2}^0 \!\!dt \int \!\!\!\!\int_{S_c}\!\!dxdyJ_z(x,y,t)E_z(x,y,t). 
\label{Q'1}
\end{equation} 
According to critical-state theory,\cite{Campbell72} during this time interval, $E_z$ is nonzero only where $J_z$ is just above $J_c$, such that $J_z$ can be replaced by $J_c$ in Eq.\ (\ref{Q'1}), but the integral is to be carried out only over those portions of the flux-penetrated cross section $S_p$, where ${\bm B}=\nabla \times {\bm A}$ is changing with time and  $E_z(x,y,t) > 0$.  Note that, if $S_I$ is chosen correctly, $E_z = 0$  throughout the entire area $S_I$ and we may therefore chose a gauge such that $A_z=0$ there.  Next, we can use Faraday's law in the form $\oint d{\bm l} \cdot {\bm E} = - \int  d{\bm S} \cdot \partial {\bm B}/\partial t = - \int  d{\bm S} \cdot \partial (\nabla \times {\bm A})/\partial t $, where the surface ${\bm S}$ is a rectangle with the sides $L_z$ parallel to the $z$ axis and the ends extending from the origin to $(x,y)$ in the flux-penetrated region.  Application of Stokes's  theorem thus yields
\begin{equation}
E_z(x,y,t)\! = -\partial A_z(x,y,t)/\partial t.
\label{Ez}
\end{equation}
Substituting this expression into Eq.\ (\ref{Q'1}), integrating over time, noting that the change in the vector potential is $A_z(x,y,0)-A_z(x,y,-T/2) =2A_z(x,y)$, we obtain 
\begin{equation}
Q' = -4 J_c \int \!\!\!\!\int_{S_p}\!\!dxdy A_z(x,y). 
\label{Q'2}
\end{equation}
(Since the current $I$ is in the positive $z$ direction at time $t = 0$,  $A_z(x,y) < 0$ outside the area $S_I$.)
The areal density of energy dissipated per unit length during one cycle therefore can be expressed as $q'(x,y) = 4J_c|A_z(x,y)|$.  For example, for a strip of rectangular cross section, we can see from the contour plot of $A_z(x,y)$ in Fig.\ 3 that the largest values of $q'(x,y)$ occur at the four corners.

To compare the losses in strips with the same critical current $I_c$ but different cross sections, it is useful to express $Q'$ in terms of a dimensionless geometry-dependent loss function $L(F)$, which is a function of $F= I/I_c$,
\begin{eqnarray}
Q'&=&\mu_0 I_c^2 L(F), 
\label{Q'} \\
L(F) &=& \frac{1}{S_c}\int \!\!\!\!\int_{S_p}\!\!dxdy\Lambda(x,y),
\label{L(F)} \\
\Lambda(x,y)&=& -4A_z(x,y)/\mu_0 I_c.
\label{Lambda}
\end{eqnarray}
Note that $\Lambda(x,y)$ is proportional to the local density of time-averaged energy dissipation.

The maximum hysteretic transport losses occur for $I=I_c$ or $F=1$, when the  flux front first touches the axis and $S_I$ shrinks to zero.  For a strip with elliptical cross section, as shown by Norris\cite{Norris70}, 
\begin{equation}
L_e(1)=1/2\pi=0.159,
\label{LeNorris}
\end{equation}
independent of the ratio $b/a$.  For a strip with rectangular cross section, as shown by Rhyner\cite{Rhyner02}, 
\begin{eqnarray}
L_r(1)&=&[6\ln 4 -7 -\tilde b^2 \ln(1+\tilde b^{-2})\nonumber \\
&-&\tilde b^{-2}\ln(1+\tilde b^2)+2 (\tilde b \tan^{-1} \tilde b^{-1} \nonumber \\
&+&\tilde b^{-1} \tan^{-1} \tilde b)]/6\pi,
\label{LrRhyner}
\end{eqnarray}
where $\tilde b = b/a$.  For a square cross section ($b=a$), $L(1) = 0.163$, and for $b/a \ll 1$, $L(1) = 0.123$, in agreement with Rhyner's results of Ref.\ \onlinecite{Rhyner02} but in disagreement with Norris's results in Table 2 of Ref.\ \onlinecite{Norris70}.
Shown in Fig.\ 6 are calculated values of $L(1)$ vs $b/a$ for various cross-sectional shapes characterized by the value of $\alpha$ in Eq.\ (\ref{Sc}).  When $b=a$, $L(1) > 0.159$ for all $\alpha>0$, with the greatest deviation occurring for a square cross section ($\alpha = \infty$).  On the other hand, when $b \ll a,$ $L(1)< 0.159$ for all $\alpha>0$, with the greatest deviation again occurring for a square cross section.  

\begin{figure}%***** Fig.6 ************************
\includegraphics[width=8cm]{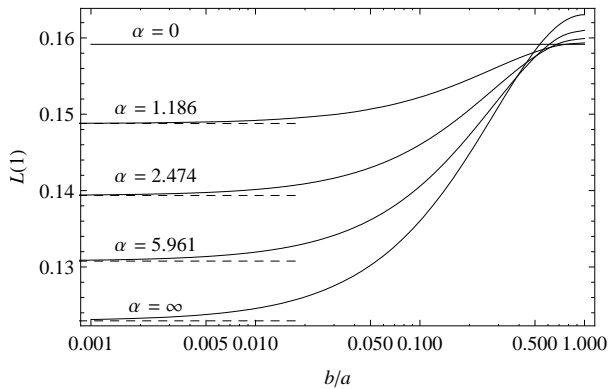}
\caption{Plots of $L(1)$ vs $b/a$ at $I=I_c$ calculated from Eqs.\ (\ref{Sc}), (\ref{Acz}), and (\ref{L(F)})-(\ref{LrRhyner}) for $\alpha = 0$ (elliptical cross section, $S_c=3.14a b$), values of $\alpha$ corresponding to intermediate shapes and cross-sectional areas (see Fig.\ 1), $\alpha= 1.186$ ($S_c= 3.36ab$), $\alpha=2.474$ ($S_c = 3.57ab$), $\alpha= 5.961$ ($S_c= 3.79ab$), and $\alpha = \infty$ (rectangular cross section, $S_c=4ab$). The cross-sectional area for $\alpha$ = 2.474 is the average of the areas of the elliptical and rectangular cross sections.  For small $b/a$ the curves asymptotically approach the corresponding thin-film limits calculated in Sec.\ \ref{lossesthin}, shown as dashed lines. }
\label{Fig06}
\end{figure} 

Another way to express the ac loss per cycle is to start from Eq.\ (\ref{Q'2}), use the symmetries $A_z(-x,y)=A_z(x,-y)=A_z(x,y)$, note that $A_z(x,y) = 0, B_x(x,y)=0,$ and $B_y(x,y) = 0$ inside $S_I$, integrate over $y$ in the first quadrant, and carry out partial integrations using $B_x = \partial A_z/\partial y$ and $B_y = -\partial A_z/\partial x$.  Equation (\ref{Q'}) then can be evaluated using
\begin{eqnarray}
L(F) &=& L_x(F)+L_y(F),
\label{Lxy} \\
L_x(F)&=&p\int_0^a dx \int_{y_l(x)}^{y_c(x)} [y_c(x)-y][-B_x(x,y)],
\label{Lx} \\
L_y(F)&=&p\int_c^a dx G(x) B_y(x,0),
\label{Ly}
\end{eqnarray}
where
\begin{eqnarray}
p&=&\frac{16}{\mu_0 J_c S_c^2},\\
y_l(x) &=& y_I(x), \; 0\le x < c,\\
&=&0,\; c\le x\le a,\\
G(x)&=&\int_x^adx'y_c(x').
\label{G}
\end{eqnarray}  
Note that $L_y$ describes the dissipation due to magnetic flux in the form of vortex or antivortex segments perpendicular to the strip transporting flux density $B_y$ in from the edges at $x =\pm a$, while $L_x$ describes the dissipation due to magnetic flux in the form of vortex or antivortex segments parallel to the strip transporting flux density $B_x$ in from the top and bottom surfaces. 

The ratio $f_y(F) = L_y(F)/L(F)$ is an increasing function of $F = I/I_c$ and a decreasing function of $\tilde b = b/a$.  Since for a fixed value of $F$, $f_y(F) \to 1$ as $b/a \to 0,$ Eq.\ (\ref{Ly}) can be used to evaluate $L(F)$ in the thin-film limit, as shown later in Fig.\ \ref{Fig09}. 

\begin{figure}%***** Fig.7 ************************
\includegraphics[width=8cm]{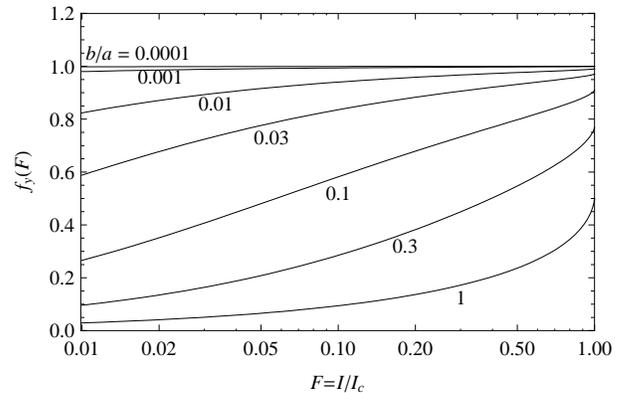}
\caption{Plots of the loss fraction $f_y(F)=L_y(F)/L(F)$ vs $F=I/I_c$ calculated from Eqs.\ (\ref{LNorrisEllipse}) and (\ref{Lyellipse})  for $b/a = 0.0001, 0.001, 0.01, 0.03, 0.1, 0.3,$ and 1 for conductors with an elliptical cross section ($\alpha = 0$).  }
\label{Fig07}
\end{figure}

\subsection{Elliptical cross section\label{losseselliptical}}

\subsubsection{$B_y$ contribution\label{lossesellipseby}}

As shown in Ref.\ \onlinecite{Norris70}, the total hysteretic ac losses of a conductor of elliptical cross section have the same remarkably simple form for all values of the ratio $b/a$,\cite{Norris70} 
\begin{equation}
L(F) = [(1 - F) \ln(1 - F) + (2 - F) F/2]/\pi,
\label{LNorrisEllipse}
\end{equation}
where $F = I/I_c$.
The fraction $f_y(F)= L_y(F)/L(F)$ easily can be evaluated numerically for a strip of elliptical cross section using  Eq.\ (\ref{Ly}) and expressions for $B_y(x,0)$ given in Ref.\ \onlinecite{Norris70},  which yield
\begin{eqnarray}
L_y(F) &=& \frac{8}{\pi^2(1-\tilde b^2)}\int_{\tilde c}^1 d\tilde x (\cos^{-1}\tilde x - \tilde x\sqrt{1-\tilde x^2}) \nonumber \\
&&\times \Big(\sqrt{\tilde x^2-\tilde c^2(1-\tilde b^2)}-\tilde b \tilde x\Big).
\label{Lyellipse}
\end{eqnarray}
For a strip with elliptical cross section $\tilde c$ and $F$ are related via $\tilde c = c/a = \sqrt{1-F}$.\cite{Norris70}  The behavior of $f_y(F)$  vs $F$ is shown in Fig.\ \ref{Fig07} for various values of $\tilde b = b/a$

\subsubsection{Behavior for small $F$ \label{lossesellipsesmallF}}

Expansion of Eq.\ (\ref{LNorrisEllipse}) in powers of $F$ yields
\begin{equation}
L(F) = \frac{F^3}{6\pi}(1+\frac{1}{2}F+\frac{3}{10}F^2+...).
\label{LNorrisSeries}
\end{equation}
The leading term in this expansion can be obtained most simply as follows.  In the Meissner state, the tangential magnetic field at the surface of an infinitely long cylinder of semimajor and semiminor axis $a$ and $b$ carrying current $I$ in the $z$ direction is\cite{Huebener72}
\begin{equation}
H_{ts}(\theta')=\frac{I}{2\pi\sqrt{(a \sin \theta')^2+(b \cos \theta')^2}},
\label{Hts}
\end{equation} 
where a point on the surface of the ellipse is described by $(x,y)=(a \cos \theta',b\sin \theta')$.  If the self-field at the surface is much larger than $H_{c1}$ (or if we assume that $H_{c1}$ is negligibly small), then according to critical-state theory, magnetic flux will penetrate to a distance $L_p(\theta') = H_{ts}(\theta')/J_c$ from the surface, assuming that $L_p(\theta')$ is much smaller than the corresponding radius of curvature of the surface, 
\begin{equation}
\rho(\theta')=[(a \sin \theta')^2+(b \cos \theta')^2]^{3/2}/ab.
\label{radcurv}
\end{equation}
When a type-II superconductor is subjected to a parallel ac field of amplitude $H_0$, the hysteretic ac loss per unit area per cycle is known to be\cite{Campbell72}
\begin{equation}
Q_A = \frac{2\mu_0 H_0^3}{3J_c}.
\label{QA}
\end{equation}
Thus the hysteretic ac loss per cycle per unit length of elliptical cylinder can be calculated to lowest order in $F$ using 
\begin{equation}
Q'=\oint dl \frac{2\mu_0 H^3_{ts}(\theta')}{3J_c}.
\label{Q'int}
\end{equation}
where $dl = \sqrt{(a \sin \theta')^2+(b \cos \theta')^2}d\theta'$ is the element of arc length.  The integral can be carried out without difficulty, yielding 
\begin{equation}
Q' =\mu_0 I_c^2 \Big(\frac{F^3}{6\pi}\Big),
\label{NorrisSmallF}
\end{equation}
where $I_c = \pi a b J_c$, which agrees with the Norris result to lowest order in $F$.  Application of the condition that $L_p(\theta') \ll \rho(\theta')$ at $\theta' =0$ is equivalent to the requirement that $F \ll (b/a)^2$ for the derivation of Eq.\ (\ref{NorrisSmallF}) to be valid.  Nevertheless, the Norris derivation, yielding the same $F^3$ behavior for small $F$, is not subject to this stringent limitation but instead yields the same result for all values of $b/a$ whenever $F \ll 1$.

\subsection{Rectangular cross section\label{lossesrectangular}}

\subsubsection{$F^3$ or $F^4$? \label{Powerlawtransition}}

Consider a strip of width $2a$ and arbitrary thickness $2b$ in the middle, as shown in Fig. 1.  According to critical-state theory, for small values of $F = I/I_c$, magnetic flux penetrates only to a small depth $L_p = H_{ts}/J_c$, where $H_{ts}$ is the self-field tangent to the surface.  The flux-penetrated cross-sectional area $S_p$ then can be thought of as a band with a geometry-dependent width that is proportional to $F$ and with a total length equal to the perimeter of the strip.   Note also that within $S_p$, $B$ varies linearly  and $A_z$ varies quadratically as a function of the distance from $S_I$, such that the integral of $A_z$ over the cross section should vary as $F^3$. Alternatively, the hysteretic ac loss per cycle per unit length can be calculated to good accuracy by using Eq.\ (\ref{QA}), replacing $H_0$ by $H_{ts}$, and integrating around the circumference of the strip.  
The hysteretic ac loss per cycle per unit length $Q'$ (and hence the loss function $L$) should always vary as $F^3$ for small $F$.
How, therefore, can we explain the Norris result for  thin strips of rectangular cross section,\cite{Norris70}
\begin{equation}
L(F) = [(1-F)\ln(1-F)+(1+F)\ln(1+F)-F^2]/\pi,
\label{LNorrisRect}
\end{equation}
which has the limiting behavior 
\begin{equation}
L(F) = F^4/6\pi
\label{LNorrisRectSmallF}
\end{equation}
for small $F$?

In short, the explanation is that although the losses  are indeed approximately proportional to $F^3$ for small $F$, the values of $F$ for which the  $F^3$ behavior holds in a thin film of rectangular cross section are very small. 
Of the terms in Eqs.\  (\ref{Lxy})-(\ref{Ly}), the $F^3$ behavior for small $F$ arises from $L_x \sim (b/a)F^3$, while $L_y \sim F^4$. Roughly speaking, $L_x$ becomes negligible relative to $L_y$ when $F \gg b/a.$  Thus, if one considers values of $F > 0.01$ for samples of rectangular cross section with $b/a \le 0.0001$, the losses due to the penetration of $B_x$  into the top and bottom surfaces, as described by $L_x$, are much smaller than the losses due to penetration of $B_y$ in from the edges, as described by $L_y$.

\subsubsection{Conformal-mapping method \label{lossesrectangularconformal}}

To describe the losses for small $F$, a calculation similar to that described in Eqs.\ (\ref{Hts})-(\ref{NorrisSmallF}) in Sec.\ \ref{lossesellipsesmallF} can be carried out for an infinitely long cylinder of rectangular cross section with width $2a$ and height $2b$.  In this case the tangential magnetic field at the surface, derived using conformal-mapping methods, is 
\begin{equation}
H_{ts}(\theta')=\frac{I}{2\pi \gamma |\sin^2 \theta' - \sin^2 \beta'|^{1/2}},
\label{Htsrect}
\end{equation} 
where $\cos \beta' = k$, $\sin \beta' = k' =\sqrt{1-k^2}$,
\begin{eqnarray}
a/\gamma &=&f = {\bm E}(k)-k'^2{\bm K}(k), \\
b/\gamma &=&f' = {\bm E}(k')-k^2{\bm K}(k'),
\end{eqnarray}
${\bm K}(k)$ is the complete elliptic integral of the first kind of  modulus $k$ and complementary modulus $k'=\sqrt{1-k^2}$,  ${\bm
E}(k)$ is the complete elliptic integral of the second
kind, and an element of arc length around the perimeter is 
$dl = \gamma |\sin^2 \theta' - \sin^2 \beta'|^{1/2}d\theta'$, starting with $(x,y) = (a,0)$, where $\theta'=0$.   At $(x,y)= (a,b)$, $\theta' = \beta'$, and at  $(x,y)= (0,b)$, $\theta' = \pi/2.$  Assuming that magnetic flux penetrates to a depth $L_p(\theta') = H_{ts}(\theta')/J_c$, using Eqs.\ (\ref{QA}) and (\ref{Q'int}), and noting that all four quadrants give equal contributions, we obtain the following integral yielding the hysteretic ac loss per cycle per unit length,
\begin{equation}
Q' = \frac{\mu_0 I^3}{3\pi^3 J_c \gamma^2}\int_0^{\pi/2}\frac{d\theta'}{|\sin^2\theta'-\sin^2\beta'|}.
\label{Q'rect}
\end{equation}
However,  at $\theta' = \beta'$, the corner $(x,y) = (a,b)$ of the rectangle, the integrand has an unphysical divergence, which needs to be cut off.
Physically, the flux fronts penetrating from the side $x = a$ and the top $y=b$ intersect at points near the corner corresponding to $\theta' = \beta' \pm \delta$, where 
\begin{equation}
\delta = \Big[\frac{9}{8kk'}\Big(\frac{L_{pm}}{\gamma}\Big)^2\Big]^{1/3} \ll 1
\end{equation}
and $L_{pm} = L_p(\beta' \pm \delta)$.  Using Eq.\ (\ref{Hts}) and eliminating $L_{pm}$ in favor of $F = I/I_c$, where $I_c = 4 a b J_c = 4 \gamma^2 f f' J_c$, we obtain
\begin{equation}
\delta = \Big(\frac{3ff'}{2\pi k k'} F\Big)^{1/2} \ll 1.
\label{deltavsF}
\end{equation}
Integrating Eq.\ (\ref{Q'rect}) over $\theta'$ from 0 to $\beta'-\delta$ to obtain the losses on the right side, replacing the integrand by its value at $\beta' \pm \delta$ over the range $\beta' - \delta < \theta' < \beta' + \delta$ to approximate the losses in the corner, integrating Eq.\ (\ref{Q'rect}) over $\theta'$ from $\beta'-\delta$ to $\pi/2$ to obtain the losses on the top, and assuming 
$\delta \ll 2kk'$
yields 
\begin{equation}
Q' =\mu_0 I_c^2 L_{rect}^{(1)}(F),
\label{RectSmallF}
\end{equation}
where 
\begin{equation}
L_{rect}^{(1)}(F) = \frac{2ff'F^3}{3\pi^3kk'}\Big[2+\ln\Big(\frac{8\pi k^3 k'^3}{3ff'F}\Big)\Big].
\label{Lrect1}
\end{equation}
The constant term within the brackets is the corner contribution, and the logarithmic term arises from two equal contributions from the length $b$ on the side and the length $a$ on the top.   Equation (\ref{Lrect1}) should be a good approximation  for small values of $F$ obeying Eq.\ (\ref{deltavsF}). 

\begin{figure}%***** Fig.8 ************************
\includegraphics[width=8cm]{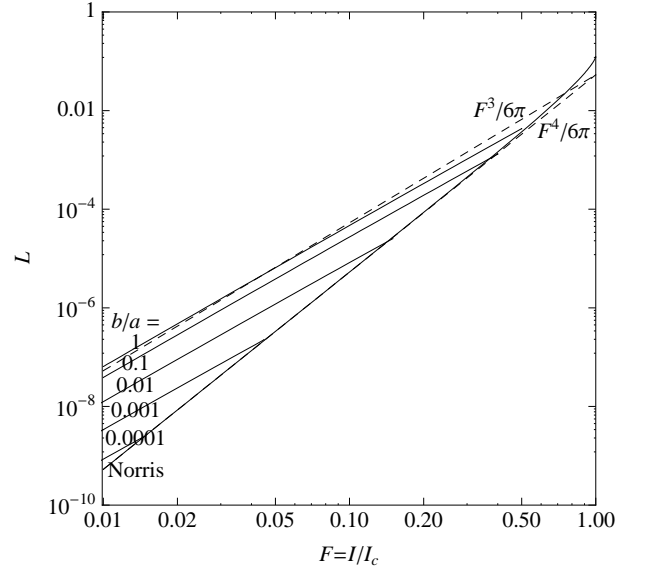}
\caption{The upper solid curves display plots of the loss function $L$ vs $F = I/I_c$ calculated from Eq.\ (\ref{Lrect1}) for a strip of rectangular cross section with relative dimensions $b/a$ = 1, 0.1, 0.01, 0.001, and 0.0001.  The bottom solid curve shows the Norris result [Eq.\ (\ref{LNorrisRect})] for a very thin strip of rectangular cross section.  For comparison, the upper dashed line shows $F^3/6\pi$ and the lower dashed curve shows $F^4/6\pi$.}
\label{Fig08}
\end{figure} 

The upper solid curves in Fig.\ \ref{Fig08} show plots of $L_{rect}^{(1)}(F)$ vs $F$ for small $F$  before the curves intersect with the Norris thin-film result for $L(F)$, Eq.\ (\ref{LNorrisRect}). For comparison, the dashed curves show the small-$F$ limits $F^3/6\pi$ for the elliptical cross section, Eq.\ (\ref{LNorrisSeries}), and $F^4/6\pi$ for thin strips of rectangular cross section, Eq.\ (\ref{LNorrisRectSmallF}).  Since the logarithmic factor in Eq.\ (\ref{Lrect1}) is slowly varying, the behavior of the hysteretic losses in a long strip of rectangular cross section must always behave approximately as $L \propto F^3$ for sufficiently small $F$, as is evident from Fig.\ \ref{Fig08}.

We can extend the above approach to somewhat larger values of $F$ as follows.  We first focus on the flux front that penetrates from the top $y=b$.  In terms of the auxiliary variable $\theta'_t$, when $\beta' \le \theta'_t \le \pi/2$, the $x$ and $y$ coordinates of a point on the top flux front in units of $a$ are  
\begin{eqnarray}
\tilde x_t(\theta'_t)&=&[E(\phi,k)-k'^2 F(\phi,k)]/f , \\
\tilde y_t(\theta'_t)&=& \frac{f'}{f}-\frac{2f'F}{\pi(\sin^2\theta'_t-\sin^2\beta')^{1/2}},
\end{eqnarray}
where $\tilde x_t(\theta'_t)=x_t(\theta'_t)/a$, $\tilde y_t(\theta'_t)=y_t(\theta'_t)/a$, $F=I/I_c$, $F(\phi,k)$
is the normal elliptic integral of the first kind\cite{Byrd71,Gradshteyn00}  of
amplitude $\phi=\arcsin(\cos \theta'_t/\cos \beta')$, modulus $k$, and complementary modulus $k'=\sqrt{1-k^2}$, and
$E(\phi,k)$ is the normal elliptic integral of the second kind.
We next characterize the flux front that penetrates from the side $x=a$.  In terms of the auxiliary variable $\theta'_s$, when $0 \le \theta'_s \le \beta'$, the $x$ and $y$ coordinates of a point on the side flux front in units of $a$ are  
\begin{eqnarray}
\tilde x_s(\theta'_s)&=&1-\frac{2f'F}{\pi(\sin^2\beta'-\sin^2\theta_s])^{1/2}}, \\
\tilde y_s(\theta'_s)&=&[E(\phi',k')-k^2 F(\phi',k')]/f,
\end{eqnarray}
where $\tilde x_s(\theta'_s)=x_s(\theta'_s)/a$, $\tilde y_s(\theta'_s)=y_s(\theta'_s)/a$, and where $\phi' =
\arcsin(\sin \theta'_s/\sin \beta')$.
The top flux front and the side flux  front intersect when $\theta'_t = \theta'_{tx}$ and $\theta'_s = \theta'_{sx}$, where $\theta'_{tx}$ and $\theta'_{sx}$ are the solutions of the following two equations:
\begin{eqnarray}
\tilde x_t(\theta'_t)&=&\tilde x_s(\theta'_s) = \tilde x_{cross},
\label{xcross}\\
\tilde y_t(\theta'_t)&=&\tilde y_s(\theta'_s) = \tilde y_{cross},
\label{ycross}
\end{eqnarray}
where $\tilde x_{cross}$ and $\tilde y_{cross}$ are the $x$ and $y$ coordinates of the crossing point in units of $a$.  Equations (\ref{xcross}) and (\ref{ycross}) always have physically reasonable solutions for sufficiently small values of $F = I/I_c$.  

The corresponding approximation to the ac loss per cycle per unit length can be obtained by integrating Eq.\ (\ref{Q'rect}) over $\theta'$ from 0 to $\theta'_{sx}$ to obtain the losses on the right side, replacing the integrand  over the range $\theta'_{sx}  < \theta' < \theta'_{tx}$ by the average   of the  values at $\theta'_{sx}$ and $\theta'_{tx}$ to approximate the losses in the corner, and integrating Eq.\ (\ref{Q'rect}) over $\theta'$ from $\theta'_{tx}$ to $\pi/2$ to obtain the losses on the top.  The result is 
\begin{equation}
Q' =\mu_0 I_c^2 L_{rect}^{(2)}(F),
\label{RectLargerF}
\end{equation}
where
\begin{eqnarray}
&&L_{rect}^{(2)}(F)= \frac{4ff'F^3}{3\pi^3}\Big[\frac{1}{kk'}\tanh^{-1}\Big(\frac{\tan \theta'_{sx}}{\tan \beta'}\Big) \nonumber \\
&&+\frac{(\theta'_{tx}-\theta'_{sx})}{2}\Big(\frac{1}{\sin^2\beta'-\sin^2\theta'_{sx}}+\frac{1}{\sin^2\theta'_{tx}-\sin^2\beta'}\Big)
\nonumber \\
&&+\frac{1}{kk'}\coth^{-1}\Big(\frac{\tan \theta'_{tx}}{\tan \beta'}\Big) \Big].
\label{Lrect2}
\end{eqnarray}
Values of the losses calculated from Eq.\ (\ref{Lrect2}) are slightly larger that those from Eq.\ (\ref{Lrect1}), but on Fig.\ \ref{Fig08} the two curves are indistinguishable.

\subsubsection{Numerical calculations \label{lossesrectangularnumerical}}

\begin{figure}%***** Fig.9 ************************
\includegraphics[width=8cm]{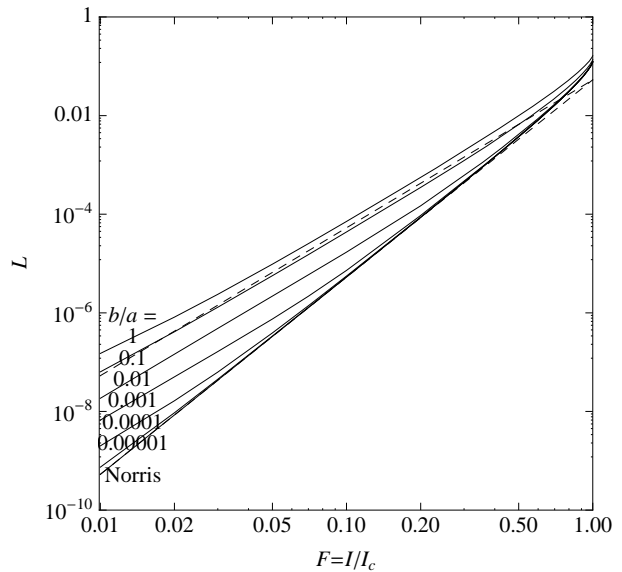}
\caption{The upper solid curves display plots of the loss function $L$ vs $F = I/I_c$ numerically calculated from Eqs.\  (\ref{Lxy})-(\ref{G}) for a strip of rectangular cross section with relative dimensions $b/a$ = 1, 0.1, 0.01, 0.001, 0.0001, and 0.00001.  The bottom solid curve shows the Norris result [Eq.\ (\ref{LNorrisRect})] for a very thin strip of rectangular cross section.  For comparison, the upper dashed line shows $F^3/6\pi$ and the lower dashed curve shows $F^4/6\pi$.}
\label{Fig09}
\end{figure} 

Shown in Fig.\ \ref{Fig09} are plots of $L(F)$ vs $F$ numerically calculated from Eqs.\  (\ref{Lxy})-(\ref{G}), where the parameters $c$, $y_0$, and $\beta$ were obtained by simultaneously solving Eq.\ (\ref{FvsSp}), $B_y(c,0)=0$, and $B_x(0,y_0)=0$.  The curves of $L(F)$ vs $F$ behave approximately as $L \propto F^3$ for all values of $b/a$ for sufficiently small $F$.  However, as $b/a$ decreases, the curves of $L(F)$ merge into the curve  Norris\cite{Norris70} obtained for a very thin strip of rectangular cross section,  Eq.\ (\ref{LNorrisRect}), which varies approximately as $F^4$.  The smaller the value of $b/a$, the smaller the value  of $F$ at which the curves merge.
For example, the numerically calculated value of $L(F)$ falls within 2\% of the Norris result, Eq.\ (\ref{LNorrisRect}), when $F>0.93$ for $b/a = 0.01$, when $F > 0.41$ for $b/a =0.001$, when $F> 0.15$ when $b/a = 0.0001$, and when $F> 0.05$ when $b/a = 0.00001$.

However, a comparison of the plots in Figs.\ \ref{Fig08} and \ref{Fig09} reveals that for small values of $F$, where the conformal-mapping result for $L(F)$ in Eq.\ (\ref{Lrect1}) is expected to be most accurate, the numerically calculated values of $L(F)$ shown in Fig.\ \ref{Fig09} are roughly a factor of two larger than the conformal-mapping result for $L(F)$ shown in Fig.\ \ref{Fig08}. 
The reason for this is that the assumed form for the flux front in Eq.\ (\ref{yI(x)}) yields a depth of magnetic-flux penetration $L_p$ near the corners at $(x,y) \approx (\pm a,\pm b)$ that is unrealistically large for small values of $F$.  Since the tangential field at the surface is approximately given by $H_{ts} \approx J_c L_p$ is correspondingly too large, and since the dissipation per cycle per unit area of surface is  approximately proportional to $H_{ts}^3$ [see Eq.\ (\ref{QA})], the dissipation per cycle per unit length, obtained by integrating around the circumference of the sample, is also  too large.   

The main story told by the results displayed in Figs.\ \ref{Fig08} and \ref{Fig09} is that in a strip of rectangular cross section the loss function $L(F)$ is approximately proportional to $F^3$ for small $F$ when $b/a$ is of the order of unity, but  as $b/a$ decreases, the $F^3$ behavior moves to smaller values of $F$, opening up a range of $F$ values for which 
$L(F)$ is approximately proportional to $F^4$.

\subsection{Thin-film limit\label{lossesthin}}

\begin{figure}%***** Fig.10 ************************
\includegraphics[width=8cm]{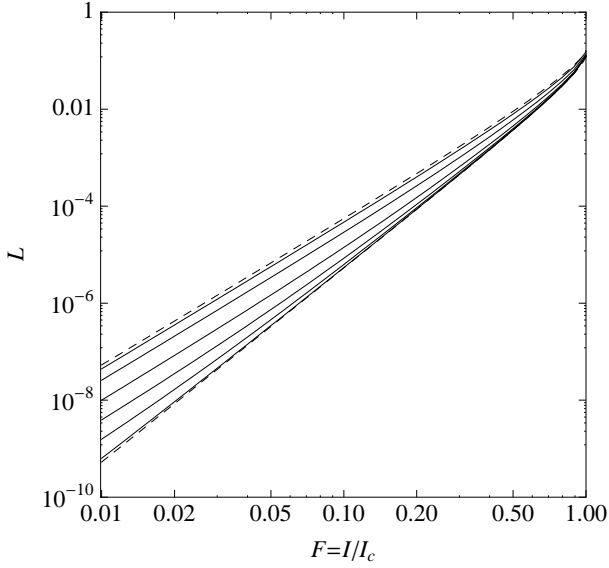}
\caption{Plots of the loss function $L$ vs $F = I/I_c$ in the thin-film limit calculated from Eq.\ (\ref{Lthin}) for strips of different cross-sectional shapes characterized by the parameter $\alpha$ = 1, 3, 10, 30, 100, and 1000 (top to bottom) in Eq.\ (\ref{y(x)}).  The top dashed curve shows the Norris result for an elliptical cross section [Eq.\ (\ref{LNorrisEllipse})], and the bottom dashed curve shows the Norris result for a thin strip of rectangular cross section [Eq.\ (\ref{LNorrisRect})].\cite{Norris70}  }
\label{Fig10}
\end{figure} 

We now calculate the hysteretic loss per cycle per unit length $Q'$ and the loss function $L$ making use of the thin-film-limit approximation that the contribution due to $L_x$  is negligible relative to that due to $L_y$.  Using Eqs.\ (\ref{complexH}) and (\ref{Ly})-(\ref{G}), we find that the loss function for a thin film described by Eq.\ (\ref{y(x)}) with the shape parameter $\alpha$ can be obtained from
\begin{equation}
L(F) = C\int_{\tilde c}^1 d\tilde x {\cal F}(\tilde x) \tilde B_y(\tilde x,0),
\label{Lthin}
\end{equation} 
where 
\begin{eqnarray}
C&=&\Big[\frac{2 \alpha }{\pi(\sqrt{1+\alpha^2}-1)}\Big]^2,\\
{\cal F}(\tilde x) &=&\frac{\sqrt{1+\alpha^2}}{\alpha}\cos^{-1}\big(\frac{\tilde x}{\sqrt{1+\alpha^2(1-\tilde x^2)}}\Big) \nonumber \\
&&-\frac{1}{\alpha}\cos^{-1}\tilde x-\tilde x \tan^{-1}(\alpha\sqrt{1-\tilde x^2}),\\
\tilde B_y(\tilde x,0)&=& \tanh^{-1}\Big[\frac{\alpha \sqrt{\tilde x^2-\tilde c^2}}{\sqrt{1+\alpha^2(1-\tilde c^2)}}\Big],
\label{Bytilde}
\end{eqnarray}
$\tilde x = x/a$, $\tilde c = c/a$, and $\tilde c$ is related to $F = I/I_c$ via Eq.\ (\ref{Fvsc}).  For $\alpha = 0$ and $\alpha = \infty$, the integral can be evaluated analytically, and the results for $L(F)$ are the same as those found by Norris,\cite{Norris70} given in Eqs.\ (\ref{LNorrisEllipse}) and (\ref{LNorrisRect}).

The six solid curves in Fig.\ \ref{Fig10} show plots of the loss function $L(F)$ numerically calculated from Eqs.\ (\ref{Lthin})-(\ref{Bytilde}) for a range of values of $\alpha$.  As expected, for small values of $\alpha$ the curves lie close to the Norris result for an elliptical cross section, and for large values of $\alpha$ the curves approach the Norris result for a rectangular cross section.

\subsection{$J_c(B)$ in the thin-film limit\label{Jc(B)}}

All the results in this paper have been carried out using the assumption that the critical current density $J_c$ is independent of the local flux density $B$.  To carry out loss calculations when $J_c$ depends strongly upon $B$ requires intensive numerical work, even in the thin-film limit, because the profiles of $B_y(x,0,t)$ and $\bar J_z(x,t)$ then must be calculated self-consistently at all times $t$ during the ac cycle.  To account approximately for the effect of self-field suppression of $J_c(B)$ upon the ac losses, we can 
make use of Eqs.\ (\ref{LNorrisEllipse}), (\ref{LNorrisRect}), and (\ref{Lthin})-(\ref{Bytilde}), which assume a constant $J_c$, by replacing $J_c$ by $J_{cI}=J_c(B_I)$, where $B_I$ is the average of $B_y(x,0)$ over the flux-penetrated band $c < x < a$ when the peak current is $I$.  From Eqs.\ (\ref{complexH})-(\ref{Finfty}) we thus obtain
\begin{eqnarray}
B_I&=& \frac{1}{a-c}\int_c^a B_y(x,0) \\
&=&\mu_0 J_{cI}b P_\alpha(\tilde c),
\label{BI}
\end{eqnarray}
where the function $P_\alpha(\tilde c)$  and its limits when $\alpha \to 0$ and $\alpha \to \infty$ are given by 
\begin{eqnarray}
P_\alpha(\tilde c)&=&\frac{1}{(1-\tilde c)\tan^{-1}\alpha}
\Big[\tanh^{-1}\sqrt{\frac{\alpha^2(1-\tilde c^2)}{1+\alpha^2(1-\tilde c^2)}}
\nonumber \\
&+&\frac{\sqrt{1+\alpha^2(1-\tilde c^2)}}{\alpha}\tanh^{-1}\sqrt{1-\tilde c^2}  \nonumber \\
&&\frac{\sqrt{1+\alpha^2}}{\alpha}\tanh^{-1}\sqrt{\frac{(1+\alpha^2)(1-\tilde c^2)}{1+\alpha^2(1-\tilde c^2)}} \Big],\\
P_0(\tilde c)&=&\frac{\sqrt{1-\tilde c^2}-\tilde c^2 \tanh^{-1}\sqrt{1-\tilde c^2} }{2(1-\tilde c)},\\
P_\infty(\tilde c)&=&\frac{2(\sqrt{1-\tilde c^2}\tanh^{-1}\sqrt{1-\tilde c^2}+\ln \tilde c ) }{\pi(1-\tilde c)},
\end{eqnarray}
and $F = I/I_c$ in Eqs.\ (\ref{Fvsc}), (\ref{F0}), and (\ref{Finfty}) must be replaced by $F_I = I/I_{cI}$, where $I_{cI} = S_c J_{cI}$.  The transport ac loss per cycle per unit length at each current-peak value $I$ is then approximated by $Q' = \mu_0 I_{cI}^2 L(F_I)$, where $L$ is given by Eq.\ (\ref{Lthin}) for $0 < \alpha < \infty$, Eq.\ (\ref{LNorrisEllipse}) for $\alpha = 0$, and Eq.\ (\ref{LNorrisRect}) for $\alpha = \infty$.  We can define $L_{eff}$ making reference to the critical current density $J_{cp}=J_c(B_p)$, where the subscript p refers to full penetration of the strip (i.e., when $\tilde c = c/a = 0$), where $B_p= \mu_0 J_{cp}b P_\alpha(0)$ and the full-penetration critical current is $I_{cp} = S_c J_{cp}$.  As a function of $F_p = I/I_{cp}$, we therefore have
\begin{equation}
L_{eff}(F_p) = \frac{Q'}{\mu_0 I_{cp}^2 }= \Big(\frac{J_{cI}}{J_{cp}}\Big)^2L\Big(F_p\frac{J_{cp}}{J_{cI}}\Big).
\label{Leff}
\end{equation} 

To proceed further, we need an explicit model for the dependence of $J_c(B)$.    Choosing the Kim model,
\begin{equation}
J_c(B) = J_c(0)/(1+B/B_0),
\end{equation}
we can solve Eq.\ (\ref{BI}) to obtain
\begin{equation}
\frac{J_{cI}}{J_{cp}}=\frac{1+\sqrt{1+4 \gamma P_\alpha(0)}}{1+\sqrt{1+4 \gamma P_\alpha(\tilde c)}},
\end{equation}
where $\tilde c$ is determined as a function of $F_p$ by numerically solving  Eq.\ (\ref{Fvsc}), (\ref{F0}), or (\ref{Finfty}) with $F$ replaced by $F_p J_{cp}/J_{cI}$.  The  dimensionless parameter  $\gamma = \mu_0 J_c(0) b/B_0$ is a measure of how strongly $J_c$ depends upon $B$.

\begin{figure}%***** Fig.11 ************************
\includegraphics[width=8cm]{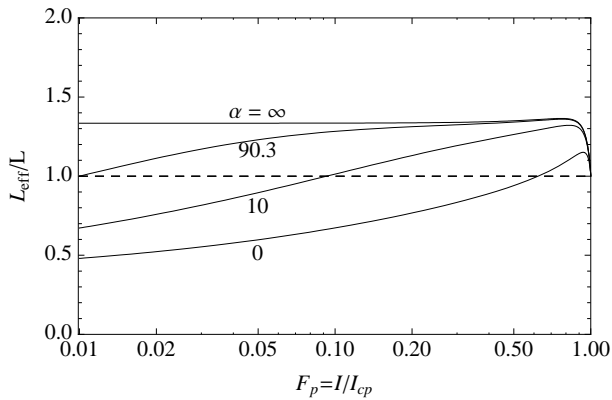}
\caption{The solid curves show plots of $L_{eff}(F_p)/L(F_p)$ for $\gamma = 10$ and values of $\alpha$ = 0 (elliptical cross section), 10, 90.3, and $\infty$ (rectangular cross section) in the thin-film limit, where $L_{eff}(F_p)$ is given by Eq.\ (\ref{Leff}) and $L(F_p)$ by Eq.\ (\ref{Lthin}).  The dashed line shows the corresponding ratios when $\gamma = 0$ and $J_c$ is independent of $B$. }
\label{Fig11}
\end{figure}

Numerical calculations for various values of $\alpha$ and $\gamma$ reveal that including the $B$ dependence of $J_c(B)$ does not have a dramatic influence upon  $L_{eff}(F_p)$.  Shown in Fig.\ \ref{Fig11} is a plot of the ratio $L_{eff}(F_p)/L(F_p)$ for $\gamma = 10$ and values of $\alpha$ = 0 (elliptical cross section), 10, 90.3, and $\infty$ (rectangular cross section) in the thin-film limit, where $L_{eff}(F_p)$ is given by Eq.\ (\ref{Leff}) and $L(F_p)$ by Eq.\ (\ref{Lthin}).  On a log-log plot the power-law dependence of $L_{eff}(F_p)$ differs only slightly from that of $L(F_p)$.

\section{Discussion\label{discussion}}

In this paper I have addressed the question of whether the ac transport losses in type-II superconducting strips should vary as $F^3$, $F^4$, or something in between, where $F = I/I_c$, $I$ is the peak alternating current, and $I_c$ is the critical current.  To account for effects of the cross-sectional shape and the thickness dependence of the strips, I did calculations assuming cross sections whose shapes are modeled by Eq.\ (\ref{y(x)}), which, by varying the shape parameter $\alpha$, describes an ellipse when $\alpha \to 0$, a rectangle when $\alpha \to \infty$, and something in between for intermediate values of $\alpha$.  

In Sec.\ \ref{fieldthick} I showed how the vector potential and magnetic field in thick films can be calculated to high accuracy using this function, and in Sec.\ \ref{fieldthin} I used the complex-magnetic-field method  to calculate the magnetic field and sheet-current density for thin films with cross sections that are elliptical, rectangular, or something in between. 
In Sec.\ \ref{losses} I discussed how to use the magnetic fields obtained as above to calculate hysteretic ac transport losses, beginning with a description of some general loss expressions in Sec.\ \ref{lossesmethods}.  

I addressed the behavior in  strips of elliptical cross section in Sec.\ \ref{losseselliptical}.  In Sec.\ \ref{lossesellipseby} I reviewed the results of Norris\cite{Norris70} and calculated  the fraction of losses attributable to the $B_y$ contribution, and  in  Sec.\ \ref{lossesellipsesmallF} I discussed the origin of the $F^3$ dependence of the ac losses for small $F$.  

For samples of rectangular cross section, discussed in Sec.\ \ref{lossesrectangular}, I concluded that the losses should always vary as $F^3$ for sufficiently small $F$.  However, I found for thin films that there is a relatively large range of values of $F$ for which the losses vary as $F^4$ and that as the film becomes very thin, the range of values of $F$ for which the $F^3$ behavior holds becomes very small.  In Sec.\ \ref{lossesrectangularconformal} I used a conformal-mapping method  to obtain approximations for the $F$ dependence of the ac losses in a rectangular strip, and in Sec.\ \ref{lossesrectangularnumerical} I presented plots of numerically calculated  ac losses for rectangular strips of various thicknesses, showing the transition between $F^3$ and $F^4$ behavior. 

In Sec.\ \ref{lossesthin} I calculated the ac losses  in very thin strips of intermediate cross section characterized by the shape parameter $\alpha$.  For the limiting cases of $\alpha = 0$ (elliptical cross section) and $\alpha = \infty$ (rectangular cross section) my results reduced to those of Norris\cite{Norris70}, but for intermediate values of $\alpha$ the calculated losses were between these two limits.

To investigate to what extent the magnetic-field dependence of the critical current density $J_c$ plays a role, in Sec.\ \ref{Jc(B)} I used the Kim model to examine the influence of $J_c(B)$ in thin films.  I found that while the $B$ dependence  does affect the magnitude of the  ac losses, it does not have a significant effect upon the $F$ dependence of the losses. 

\begin{acknowledgments} I thank A. P. Malozemoff for posing questions that
stimulated my work on this problem, and I thank V. G. Kogan for helpful advice.  Work at the Ames Laboratory was
supported by the Department of Energy - Basic Energy Sciences under Contract
No. DE-AC02-07CH11358.
\end{acknowledgments}

\appendix 
\begin{widetext} 
\section{Elliptical cross section}

The vector potential generated by a uniform current density $J_c$ flowing in the $z$ direction through a cylinder of elliptical cross section centered on the $z$ axis, having semimajor axis $a$ along the $x$ axis and semiminor axis $b$ along the $y$ axis is ${\bm A}_{ce}(x,y) = A_{cze}(x,y) \hat z$ [Eq.\ (\ref{Acz})], where 
\begin{equation}
A_{cze}(x,y)=-\frac{\mu_0 J_c (b x^2+a y^2)}{2(a+b)}
\label{insideellipse}
\end{equation}
on or inside the ellipse $(x/a)^2+(y/b)^2 = 1$, and 
\begin{eqnarray}
A_{cze}(x,y)&=&-\Re\frac{\mu_0 J_c a b}{2(a^2-b^2)}\Big[\zeta(\zeta-\sqrt{\zeta^2-a^2+b^2})
+(a^2-b^2)\ln\Big(\frac{\zeta+\sqrt{\zeta^2-a^2+b^2}}{a+b}\Big)\Big], \; b \ne a, 
\label{outsideellipse}\\
&=&-\Re\frac{\mu_0 J_c a^2}{2}\Big[\frac{1}{2}
+\ln\Big(\frac{\zeta}{a}\Big)\Big], \; b = a,
\label{outsidecircle}
\end{eqnarray}
on or outside the ellipse, where $\zeta = x+iy$ and $\Re$ denotes the real part.

Analytic expressions for the $x$ and $y$ components of the corresponding flux density ${\bm B}_{ce}(x,y)=\nabla \times {\bm A}_{ce}(x,y)$ are given in Ref.\ \onlinecite{Norris70}.

\section{Rectangular cross section}

The vector potential generated by a uniform current density $J_c$ flowing in the $z$ direction through a cylinder of rectangular cross section centered on the $z$ axis, having width $2a$ along the $x$ axis and height $2b$ along the $y$ axis is ${\bm A}_{cr}(x,y) = A_{czr}(x,y) \hat z$ [Eq.\ (\ref{Acz})], where

\begin{eqnarray}
A_{czr}(x,y)& =& \frac{\mu_0 J_c}{4 \pi}\Big\{(x-a)^2\Big[\tan^{-1}\Big(\frac{y+b}{x-a}\Big)-\tan^{-1}\Big(\frac{y-b}{x-a}\Big)\Big]
+(x+a)^2\Big[\tan^{-1}\Big(\frac{y-b}{x+a}\Big)-\tan^{-1}\Big(\frac{y+b}{x+a}\Big)\Big] \nonumber \\
&+&(y-b)^2\Big[\tan^{-1}\Big(\frac{x+a}{y-b}\Big)-\tan^{-1}\Big(\frac{x-a}{y-b}\Big)\Big]
+(y+b)^2\Big[\tan^{-1}\Big(\frac{x-a}{y+b}\Big)-\tan^{-1}\Big(\frac{x+a}{y+b}\Big)\Big] \nonumber \\
&+&(x-a)(y+b)\ln[(x-a)^2+(y+b)^2]-(x-a)(y-b)\ln[(x-a)^2+(y-b)^2]\nonumber \\
&+&(x+a)(y-b)\ln[(x+a)^2+(y-b)^2]-(x+a)(y+b)\ln[(x+a)^2+(y+b)^2]\nonumber \\
&+&4a^2\tan^{-1}\Big(\frac{b}{a}\Big)+4b^2\tan^{-1}\Big(\frac{a}{b}\Big)
+4ab\ln(a^2+b^2).\Big\}
\label{Aczr}
\end{eqnarray}

The $x$ and $y$ components of the corresponding flux density ${\bm B}_{cr}(x,y)=\nabla \times {\bm A}_{cr}(x,y)$ are 

\begin{eqnarray}
B_{cxr}(x,y)&= &
\frac{\mu_0 J_c}{4 \pi}\Big\{2(y-b)[\arctan(\frac{x+a}{y-b})-\arctan(\frac{x-a}{y-b})]
+2(y+b)[\arctan(\frac{x-a}{y+b})-\arctan(\frac{x+a}{y+b})]
\nonumber \\
&&+(x+a)\ln[\frac{(x+a)^2+(y-b)^2}{(x+a)^2+(y+b)^2}]
+(x-a)\ln[\frac{(x-a)^2+(y+b)^2}{(x-a)^2+(y-b)^2}]\Big\},
\label{Bcxr}
\end{eqnarray}

\begin{eqnarray}
B_{cyr}(x,y)&=&
\frac{\mu_0 J_c}{4 \pi}\Big\{2(x-a)[\arctan(\frac{y-b}{x-a})-\arctan(\frac{y+b}{x-a})]
+2(x+a)[\arctan(\frac{y+b}{x+a})-\arctan(\frac{y-b}{x+a})]\nonumber\\
&&+(y-b)\ln[\frac{(x-a)^2+(y-b)^2}{(x+a)^2+(y-b)^2}]
+(y+b)\ln[\frac{(x+a)^2+(y+b)^2}{(x-a)^2+(y+b)^2}]\Big\}.
\label{Bcyr}
\end{eqnarray}

\section{Numerical calculation of ${\bm B}_I(x,y)$}

The auxiliary magnetic induction  ${\bm B}_I(x,y)  = \nabla \times {\bm A}_{I}(x,y)$ [see Eq.\ (\ref{AIz})] is generated by a uniform current density $J_z = -J_c$  flowing only in the cross section $S_I$.  The $x$ and $y$ components of ${\bm B}_I(x,y)$ readily can be    calculated numerically from the following one-dimensional integrals 
\begin{equation}
B_{Ix}(x,y) = \frac{\mu_0 J_c}{4 \pi}\int_{-c}^c \ln\Big\{
\frac{(u-x)^2+[y_I(u)+y]^2}{(u-x)^2+[y_I(u)-y]^2}\Big\}du,
\label{BIx}
\end{equation}
\begin{equation}
B_{Iy}(x,y) = \frac{\mu_0 J_c}{4 \pi}\int_{-y_0}^{y_0} \ln\Big\{
\frac{[x_I(v)-x]^2+(v-y)^2}{[x_I(v)+x]^2+(v-y)^2}\Big\}dv,
\label{BIy}
\end{equation}
where the functions $y_I(x)$ and $x_I(y)$ are given in Eqs.\ (\ref{yI(x)}) and (\ref{xI(y)}).

\end{widetext}

\end{document}